\documentstyle[12pt]{article}
\def\ftoday{{\sl  \number\day \space\ifcase\month
\or Janvier\or F\'evrier\or Mars\or avril\or Mai
\or Juin\or Juillet\or Ao\^ut\or Septembre\or Octobre
\or Novembre \or D\'ecembre\fi
\space  \number\year}}
\newcommand{\journal}[4]{{\em #1~}#2\,(19#3)\,#4;}

\newcommand{\hpa}{\journal {Helv. Phys. Acta}}

\newcommand{\cmp}{\journal {Commun. Math. Phys.}}

\newcommand{\np}{\journal {Nucl. Phys.}}
\newcommand{\pl}{\journal {Phys. Lett.}}

\newcommand{\nc}{\journal {Nuovo Cimento}}

\makeatletter
\@addtoreset{equation}{section}
\makeatother

\newcommand{\es}{\\[3mm]}

\renewcommand{\a}{\alpha}
\renewcommand{\b}{\beta}
\newcommand{\g}{\gamma}           
\renewcommand{\d}{\delta}         \newcommand{\D}{\Delta}
\newcommand{\e}{\varepsilon}

\newcommand{\m}{\mu}
\newcommand{\n}{\nu}

\newcommand{\r}{\rho}
\newcommand{\s}{\sigma}           
\newcommand{\th}{\theta}         
\newcommand{\f}{{\phi}}           
\newcommand{\vf}{{\varphi}}

\newcommand{\PP}{{\cal P}}

\newcommand{\SS}{{\cal S}}

\newcommand{\WW}{{\cal W}}

\newcommand{\Sla}{\raise.15ex\hbox{$/$}\kern -.70em D}

\newcommand{\lp}{\left(}\newcommand{\rp}{\right)}
\newcommand{\lc}{\left[}\newcommand{\rc}{\right]}
\newcommand{\lac}{\left\{}\newcommand{\rac}{\right\}}

\newcommand{\complex}{{\kern .1em {\raise .47ex
\hbox {$\scriptscriptstyle |$}}
    \kern -.4em {\rm C}}}
\newcommand{\real}{{{\rm I} \kern -.19em {\rm R}}}
\newcommand{\rational}{{\kern .1em {\raise .47ex
\hbox{$\scripscriptstyle |$}}
    \kern -.35em {\rm Q}}}
\renewcommand{\natural}{{\vrule height 1.6ex width
.05em depth 0ex \kern -.35em {\rm N}}}

\newcommand{\tr}{{\rm {Tr} \,}}
\newcommand{\half}{\frac 1 2}
\newcommand{\pa}{\partial}

\newcommand{\dfud}[2]{{\displaystyle{\frac{\delta #1}{\delta #2}}}}

\newcommand{\dsum}[2]{\displaystyle{\sum_{#1}^{#2}}}
\newcommand{\dint}{\displaystyle{\int}}

\newcommand{\ie}{{{\em i.e.},\ }}

\newcommand{\sla}{\raise.15ex\hbox{$/$}\kern -.57em}
\newcommand{\twiddle}{\lower.9ex\rlap{$\kern -.1em\scriptstyle\sim$}}

\newcommand{\vev}[1]{\left\langle {#1}\right\rangle}
\newcommand{\equ}[1]{(\ref{#1})}
\newcommand{\eq}{\begin{equation}}
\newcommand{\eqn}[1]{\label{#1}\end{equation}}
\newcommand{\eea}{\end{eqnarray}}
\newcommand{\eqa}{\begin{eqnarray}}
\newcommand{\eqan}[1]{\label{#1}\end{eqnarray}}
\newcommand{\ba}{\begin{array}}
\newcommand{\ea}{\end{array}}
\newcommand{\eqac}{\begin{equation}\begin{array}{rcl}}
\newcommand{\eqacn}[1]{\end{array}\label{#1}\end{equation}}

\begin{document}
\newcommand{\pic}{$\spadesuit\spadesuit$}
\newcommand{\zc}{Z_{\rm c}}
\newcommand{\bs}{\bar s}
\newcommand{\bv}{\bar \nu}
\newcommand{\xt}{x^{\rm tr}}
\newcommand{\yt}{y^{\rm tr}}
\newcommand{\zt}{z^{\rm tr}}
{\hfill {\bf UGVA---DPT 1993/05--817}}

{\hfill {\bf TUW 93-10}} \vspace{25mm}

\noindent{\Large{\bf Symmetries of the Chern-Simons Theory \\[2mm]
                     in the Axial Gauge}}\vspace{13mm}

\noindent {\large By A. Brandhuber,  M. Langer, M. Schweda,  S.P.
                  Sorella$^{1,2}$}\vspace{3mm}

\noindent Institut f\"ur Theoretische Physik,
          Technische Universit\"at Wien \\
          Wiedner Hauptstra\ss e 8-10,
          A-1040 Wien (Austria)
\vspace{5mm}

\noindent {\large S. Emery and  O. Piguet$^1$}\vspace{3mm}

\noindent D\'epartement de Physique Th\'eorique,
                                  Universit\'e de Gen\`eve\\
          24, quai Ernest Ansermet,
          CH -- 1211 Gen\`eve 4 (Switzerland)
\vspace{24mm}

\footnotetext[1]{Supported in part
                 by the Swiss National Science Foundation.}
\footnotetext[2]{ Supported in part
     by the ''Fonds zur F\"orderung der Wissenschaftlichen Forschung'',
     M008-Lise Meitner Fellowship.}

\noindent {\small {\em Abstract.}
The Green functions of the Chern-Simons theory quantized in the axial
gauge are shown to be calculable as the
unique, exact solution of the  Ward
identities which express the invariance of the theory under the topological
supersymmetry of Delduc, Gieres and Sorella.
}\vspace{6mm}
\vfill\pagebreak

\section{Introduction}
It was shown in a previous paper~\cite{1} that the Chern-Simons
model~\cite{schwarz,witten,froelich-king}, quantized
in the axial gauge\footnote{See also~\cite{gajdosik} for more general
noncovariant gauges.}, obeys
the topological supersymmetry which was known to hold in the Landau
(covariant) gauge~\cite{brt,dgs}. This supersymmetry, whose generators form a
three-vector and whose anticommutator with the BRS generator yields the
translations, has been shown to be at the origin of the utraviolet
finiteness of the theory~\cite{dlps,bps}.
 Another important feature of topological supersymmetry, which makes
it physically relevant\footnote{ Concerning the physical irrelevance
of topological supersymmetry we
disagree with the authors of Ref.~\cite{gmrr}},
is its role in the construction of observables~\cite{gms}.

 The aim of the present work is to
examine further the relevance and the consequences of supersymmetry.
This will be done in the axial gauge,
since in this particular gauge all
calculations can be performed rather
explicitly~\cite{froelich-king,emery-piguet,6,7,kapustin}.
 The extension
of our discussion to other gauge choices remains to be done.

We will show that the  Ward identities defining the theory,
 namely the Ward identities for gauge invariance and for
supersymmetry,
allow to compute exactly all the Green functions, without the need
of an action principle and of the usual Feynman graph expansion
derived from it. Our main result is that the solution of the Ward
identities is explicit and unique.  This solution turns out
to coincide with the expression which
one would have obtained from the mentionned Feynman graph expansion,
only tree graphs contributing to it, an the principal value
prescription~\cite{kummer} being chosen for the free propagators.

The importance of the
topological supersymmetry is stressed by the outcome that the latter,
together with the axial gauge condition, essentially suffices to
determine  the
theory. All Green functions of the gauge and ghost fields indeed
are fixed by
the supersymmetry Ward identities, without demanding gauge or BRS
invariance. They moreover coincide with the ones which would follow from
the {\em BRS invariant} action. The {\em resulting} BRS invariance in
turn fixes the Green functions involving the Lagrange multiplier
field\footnote{The Lagrange multiplier fields are
used  for the implementation of the gauge condition.}
solely, which are the only ones not determined by supersymmetry.

Another intriguing point of the Chern-Simons model in the axial gauge is
the existence of a very large algebra of symmetries. It was already
shown in Ref.~\cite{1} that an anti-BRS invariance~\cite{2} and its associated
supersymmetry hold. But these new invariances do not make a closed
algebra with the ones already known. They belong to an algebra which is
shown in App.~\ref{algebra}.
\section{Chern-Simons theory in the axial gauge}
The action of the Chern-Simons model in the axial gauge
reads\footnote{{\bf Conventions}: $\m,\n,\cdots = 1,2,3\ $,
$g_{\m\n}= {\rm diag}(1,-1,-1)\ $, $\e^{\m\n\r}=\e_{\m\n\r}=\e^{[\m\n\r]}$,
$\e_{123}=1$.}
\eq\ba{rl}
\Sigma_{\rm CS}=&-\frac{1}{2}\dint d^3x \epsilon^{\mu \nu \rho}\tr(A_\mu
\partial_\nu A_\rho +\frac{2}{3} g A_\mu A_\nu A_\rho)   \es
            &+\dint d^3x \tr(d n^\mu A_\mu+b n^\mu D_\mu c),
\ea\eqn{action}
with $D_\mu\cdot=\partial_\mu \cdot +g[A_\mu,\cdot]$ for the
covariant derivative. The gauge group is
chosen to be simple, all fields belong to the adjoint representation and
are written as Lie algebra
matrices $\varphi(x)=\varphi^a(x) \tau_a$, with
\[
[\tau_a,\tau_b]= f_{ab}^c \tau_c,\qquad \tr(\tau_a
\tau_b)=\delta_{ab} .
\]
The canonical dimensions and ghost numbers of the fields are given in
Table~\ref{dim-fp}.
\begin{table}[hbt]
\centering
\begin{tabular}
{|l|              r|     r|     r|    r|    } \hline
                &$A$   &$d$   &$b$  &$c$    \\ \hline
Dimension       &$1$   &$2$   &$2$  &$0$      \\ \hline
Ghost number    &$0$   &$0$  &$-1$  &$1$     \\ \hline
\end{tabular}
\caption[t1]{Dimensions and ghost numbers.}
\label{dim-fp}
\end{table}
\section{Symmetries and Ward identities}
The action~\equ{action}
is invariant~\cite{1}
under the BRS and anti-BRS transformation $s$ and $\bs$:
\eq\ba{ll}
sA_\m =- D_\mu c, \qquad    & \bar{s} A_\mu=-D_\mu b ,\es
s b = d ,                   &\bar{s} b = g b^2,\es
s c = g c^2,                &\bar{s} c =d ,\es
s d = 0 ,                   &\bar{s} d= 0,
\ea\eqn{brs}
as well as under the vector supersymmetries $\n_\m$ and $\bv_\m$:
\eq\ba{ll}
\n_\r  A_\mu =\epsilon_{\r  \mu \nu}n^\nu b, \qquad
      &\bar{\nu}_\r  A_\mu=\epsilon_{\r  \mu \nu}n^\nu c,\es
\n_\r  b =0,             &\bar{\nu}_\r  b =-A_\r ,\es
\n_\r  c = -A_\r,        &\bar{\nu}_\r  c = 0,\es
\n_\r  d =\partial_\r b, &\bar{\nu}_\r  d = \partial_\r  c .
\ea\eqn{susy}
The BRS transformations $s$ and the supersymmetry transformations $\n_\m$
form an algebra which closes on-shell~\cite{dgs}:
\eq
s^2=\{\n_\m ,\n_\n\}=0,\quad \{s,\n_\m \}=\partial_\m
           + {\rm Eq.\ of\ motion}.
\eqn{s-v-alg}

There is a similar algebra for $\bs$ and $\bv_\m$.
However, if one wants to consider the whole set of
transformations  \equ{brs} and \equ{susy}, it
must be completed in order to
form a closed algebra. This is done in App.~\ref{algebra}.
We keep now only
the BRS and supersymmetry transformations $s$ and $\n_\m$. We shall also
assume that the theory is scale invariant, as it is the case for
the classical theory.

The BRS invariance of the theory can be expressed, formally, by the
functional identity
\eq
\tr\dint d^3x \left(
 - J^\m  [D_\m c]\cdot\zc   - g J_c [c^2]\cdot\zc
                - J_b \dfud{\zc}{J_d}   \right)  = 0.
\eqn{slavnov}
Here $\zc(J^\m,J_b,J_c,J_d)$ is the generating functional of
the connected Green
functions, $J^\m$, $J_d$, $J_b$ and $J_c$ denoting the sources of the
fields $A_\m$, $d$, $b$ and $c$, respectively. We have used the
notation
\[[O]\cdot\zc(J^\m,J_b,J_c,J_d)\]
for the generating functional of
the connected Green functions with the insertion of the
local field polynomial
operator $O$. Usually, such insertions must be renormalized, their
renormalization is controlled by coupling them to external fields and
the identity \equ{slavnov} becomes the Slavnov
identity~\cite{piguet-rouet}.
We shall however see below that, in the axial gauge which we will choose to
work in, these insertions are trivial and thus the Slavnov identity is
replaced by a local gauge Ward identity.

The axial gauge is defined by the {\em gauge condition}
\eq
n^\m \dfud{\zc}{J^\m} + J_d =  0.
\eqn{gauge-cond}
This gauge choice breaks Poincar\'e invariance, but the theory remains
invariant under the {\em transverse Poincar\'e
group}, i.e. under the Poincar\'e transformations which leave the gauge vector
$n$ unchanged.

The invariance under the supersymmetry transformations $\n_\r$
given in \equ{susy}
 leads to the {\em supersymmetry Ward identity}
\eq
\tr\dint d^3x \lp
 J^\m\e_{\r\m\n} n^\n\dfud{}{J_b}
+ J_c\dfud{}{J^\r} + J_d\pa_\r \dfud{}{J_b} \rp \zc = 0.
\eqn{susy-wi}

The projection of the supersymmetry Ward identity along the gauge vector
$n$,
\eq
\tr\dint d^3x J_d \lp -J_c + n^\m\pa_\m \dfud{\zc}{J_b} \rp = 0,
\eqn{susy-wi-long}
can be put in a more convenient form. Calling $X$ the term
between the parenthesis we see that locality, scale
invariance and ghost number conservation imply that $X$ is a local
polynomial in the sources $J$ and the functional derivatives $\d/\d J$,
of dimension 3 and ghost number $-1$.
Its most general form, compatible with transverse Poincar\'e invariance
and with the gauge condition \equ{gauge-cond} taken into account,
reads -- in component form:
\[
X_a = x\, J_{c^a} + y\, n^\m\pa_\m \dfud{\zc}{J_{b^a}}
                      + z_{abc}\,J_d^b\dfud{\zc}{J_{b^c}} ,
\]
where $x$, $y$ and $z_{abc}$ are constants,
the latter being a tensor invariant under the
gauge group. Equation~\equ{susy-wi-long}, which reads
\[
\dint d^3x J_{d^a} X_a = 0,
\]
holds if and only if $x=y=0$ and $z_{abc}$ is
 antisymmetric in $a$
and $b$, \ie is proportional to the structure constants $f_{[abc]}$.
Thus, coming back to the matrix notation, we see that
$X$ is proportional to the commutator of $J_d$ with the functional
derivative  of $\zc$ with respect to $J_b$. Calling $g$ the
proportionality factor, we get in this way the
local {\em  antighost equation}
\eq
 - J_c  +
\lp n^\m\pa_\m\dfud{}{J_b} - g \lc J_d,\dfud{}{J_b} \rc \rp \zc = 0.
\eqn{antighost-eq}

We have thus seen that the gauge condition together with the
$n$-component \equ{susy-wi-long} of the supersymmetry Ward
identity imply the local antighost equation.
The converse statement being obvious,
it follows that, under the axial gauge condition,
the local antighost equation is
indeed equivalent to the $n$-component of the supersymmetry.

 As in any gauge theory with a linear gauge condition there is a
{\em ghost equation}. This follows from the Slavnov identity
\equ{slavnov},
differentiated with respect to the source $J^3$,
and from the gauge condition \equ{gauge-cond}. The ghost equation reads
\eq
 - J_b  +
\lp n^\m\pa_\m\dfud{}{J_c} - g \lc J_d,\dfud{}{J_c} \rc \rp \zc = 0.
\eqn{ghost-eq}

 The ghost equation \equ{ghost-eq}
and the antighost equation \equ{antighost-eq}
express the ''freedom'' of the ghosts in the axial
gauge~\cite{kummer}: they couple only to the external source $J_d$,
\ie to the $n$-component of the gauge field. Their  effect is
to factorize out the contributions of the
ghost field $c$ to the composite fields appearing in the BRS Ward
identity \equ{slavnov}.  We shall thus replace the latter by the
local {\em gauge Ward identity}:
\eq\ba{l}
-\pa_\m J^\m + \lp g\lc J^\m ,\dfud{}{J^\m}\rc
  + g\lc J_d, \dfud{}{J_d}\rc + g\lac J_b, \dfud{}{J_b}\rac \right.\es
\left.\qquad  + g\lac J_c, \dfud{}{J_c}\rac -
                   n^\m \pa_\m \dfud{}{J_d}\rp\zc = 0  .
\ea\eqn{gauge-wi}
\section{Consequences of supersymmetry}
Let us show that, taken together with the axial gauge condition
\equ{gauge-cond} and the requirement of scale invariance,
the supersymmetry Ward identities \equ{susy-wi} determine all the
Green functions of the theory except those containing only the Lagrange
multiplier field $d$.

Without loss of generality we can choose the vector $n$ defining the
axial gauge as
\eq
(n^\m) = (0,0,1).
\eqn{gauge-vector}
The coordinates transverse to $n$ will be denoted by
\eq
\xt = (x^i,\ i=1,2).
\eqn{trans-coord}

Since the supersymmetry Ward identity \equ{susy-wi} for $\r=3$ is
equivalent to the antighost equation \equ{antighost-eq}, we shall solve the
latter:
\eq
 - J_c  +
\lp \pa_3\dfud{}{J_b} - g \lc J_d,\dfud{}{J_b} \rc \rp \zc = 0,
\eqn{antighost-eq-i}
and then
the transverse components of the supersymmetry Ward identity\footnote{with
$\e_{ij}=\e_{[ij]}$, $\e_{12}=1$}:
\eq
\tr\dint d^3x \lp
 J^j\e_{ij}\dfud{}{J_b}
+ J_c\dfud{}{J^i} + J_d\pa_i \dfud{}{J_b} \rp \zc = 0.
\eqn{susy-wi-tr}
\subsection*{Gauge condition:}
We begin by looking at the gauge condition
\equ{gauge-cond}. It implies the vanishing of all
connected Green functions involving the component $A_3$ of the
gauge field:
\eq
\vev{A^{a}_3(x)\vf_1(x_1)\cdots\vf_N(x_N)}=0,\quad
\forall \ {\rm fields}\ \vf_k(x_k) ,
\eqn{g-conditions}
with one exception:
\eq
\vev{A^{a}_3(x)d^b({y})} = -\d^{ab}\d^3(x-y).
\eqn{g-cond-exc}
\subsection*{Antighost equation:}
The antighost equation \equ{antighost-eq} gives the following equations for the
connected Green functions involving  one pair of ghost fields:
\eq
\pa_{x^3} \vev{ b^a(x) c^b(y) } = -\d^{ab} \d^3(x-y),
\eqn{gh-equation}
and
\eq\ba{l}
\pa_{x^3} \vev{  b^a(x) c^b(y) d^{c_1}(z_1) \cdots d^{c_n}(z_n)
  A_{i_1}^{d_1}(t_1) \cdots  A_{i_m}^{d_m}(t_m)  } \es
= g \dsum{k=1}{n} f_{ac_ke} \d^3(x-z_k) \es
\qquad\vev{  b^e(z_k) c^b(y)
  d^{c_1}(z_1) \cdots \widehat{d^{c_k}}(z_k) \cdots d^{c_n}(z_n)
  A_{i_1}^{d_1}(t_1) \cdots  A_{i_m}^{d_m}(t_m)  } \es
{\rm for}\quad (n,m) \not= (0,0)    ,
\ea\eqn{gh-equations}
where $\widehat{X}$ means the omission of the argument $X$. The
right-hand-side of \equ{gh-equations}
of course vanishes for $(n,m)=(0,m),\ m\not=0$.

The general solution of the differential equation  \equ{gh-equation} reads
\eq
\vev{ b^a(x) c^b(y) } = -\d^{ab} [\th(x^3-y^3)+\a] \d^2(\xt-\yt),
\eqn{<bc>}
where $\th$ is the step function
\eq
\th(u)= \cases{1  &if\ $u>0$,\cr
               0  &if\ $u<0$, \cr}
\eqn{step-function}
and $\a$ is an integration constant. The term proportional to $\a$ represents
the general solution of the homogeneous equation, compatible with
transverse Poincar\'e invariance and scale invariance\footnote{Scale
invariance excludes a solution of the type $1/(\xt-\yt)^2$. Indeed, due to
its short distance singularity,
the latter expression is not a well defined distribution. To give it a
meaning would need the introduction of a UV subtraction point, \ie of a
dimensionful parameter which would break scale invariance.}.

Using the same arguments we can write the
general solution of the system of equations \equ{gh-equations} as
\eq\ba{l}
\vev{  b^a(x) c^b(y) d^{c_1}(z_1) \cdots d^{c_n}(z_n)
  A_{i_1}^{d_1}(t_1) \cdots  A_{i_m}^{d_m}(t_m)  } \es
= g \dsum{k=1}{n} f_{ac_ke} [\th(x^3-z_k^3)+\a^{(n,m)}] \d^2(\xt-\zt_k)\es
\qquad \vev{  b^e(z_k) c^b(y)
  d^{c_1}(z_1) \cdots \widehat{d^{c_k}}(z_k) \cdots d^{c_n}(z_n)
  A_{i_1}^{d_1}(t_1) \cdots  A_{i_m}^{d_m}(t_m)  } \es
{\rm for}\quad (n,m) \not= (0,0).
\ea\eqn{<bcda>}
As in \equ{gh-equations} the right-hand-side vanishes for $n=0$.
The  integration constants
$\a^{(n,m)}$  depend only on the numbers $n$ and $m$,  and not on $k$,
due to the Bose symmetry of the field $d$.

Equations~\equ{<bcda>} build a recurrence on the number of fields $d$.
They imply that from the
connected Green functions with two ghost fields, those containing the
field $A$ vanish,
\eq
\vev{  b c (d)^n (A)^m } = 0\qquad {\rm for}\quad m \not= 0 ,
\eqn{<bcdaa>}
where an obvious shortened notation has been used.
On the other hand the Green functions not containing
the field $A$ are completely determined
by the ghost propagator \equ{<bc>} -- up to the integration constants
$\a^{(n)}\equiv\a^{(n,0)}$ -- through the recurrence relations
 \equ{<bcda>} taken for $m=0$:
\eq\ba{l}
\vev{  b^a(x) c^b(y) d^{c_1}(z_1) \cdots d^{c_n}(z_n)  } \es
= g \dsum{k=1}{n} f_{ac_ke} [\th(x^3-z_k^3)+\a^{(n)}]
                                               \d^2(\xt-\zt_k)\es
\qquad \vev{  b^e(z_k) c^b(y)
  d^{c_1}(z_1) \cdots \widehat{d^{c_k}}(z_k) \cdots d^{c_n}(z_n) } \es
{\rm for}\quad n\ge1.
\ea\eqn{<bcd>}

A very similar argument shows that the connected Green functions
involving more than one pair of ghosts all vanish. One indeed sees from
the antighost equation \equ{antighost-eq} that the Green functions of the type
$\vev{(b)^p (c)^p}$ vanish for $p>1$. The
recurrence relations generalizing \equ{<bcda>} then imply the result
\eq
\vev{(b)^p (c)^p (d)^n (A)^m} = 0 ,\quad {\rm for}\ p>1,
\eqn{more-bcda}
whith the same shortened notation as in \equ{<bcdaa>}.
\subsection*{Transverse supersymmetry:}
The transverse supersymmetry Ward identity \equ{susy-wi-tr} yields, for
the two-point functions:
\eq\ba{l}
\vev{A^a_i(x)A^b_{j}(y)} +
   \e_{ij3}\vev{b^a(x)c^b(y)} = 0,\es
\vev{d^a(x)A^b_i(y)} - \pa_{x^i}\vev{b^a(x)c^b(y)} = 0.
\ea\eqn{susy-conditions-2}
With the result \equ{<bc>}, this gives
\eq
\vev{A^a_i(x)A^b_j(y)} =
   \e_{ij}  \d^{ab} \lc \th(x^3-y^3)+\a \rc \d^2(\xt-\yt),\es
\eqn{<aa>}
\eq
\vev{d^a(x)A^b_i(y)} =
   -  \d^{ab} \lc \th(x^3-y^3)+\a \rc \pa_{x^i} \d^2(\xt-\yt).
\eqn{<da>}
The integration constant $\a$ is now fixed to the value
\eq
\a=-\half
\eqn{kdemi}
by the Bose symmetry condition on the propagator $\vev{A_i A_j}$.
This result
corresponds to the Cauchy principal value prescription~\cite{kummer}
for the
propagator in momentum space. Indeed the Fourier transform of
$\lc\th(u)-\half\rc\d^2(\xt)$ is equal to
\[
{\rm vp}\,\frac{i}{p_3} =  {\rm vp}\,\frac{i}{n^\m p_\m}
\]

For the higher point connected Green functions the tranverse
supersymmetry Ward identity \equ{susy-wi-tr} gives the relations
\eq\ba{l}
\vev{ A_i^b(y) d^{c_1}(z_1) \cdots d^{c_n}(z_n)
  A_{i_1}^{d_1}(t_1) \cdots  A_{i_m}^{d_m}(t_m)  } \es
=  \dsum{k=1}{n} \pa_{z^i_k} \vev{  b^{c_k}(z_k) c^b(y)
  d^{c_1}(z_1) \cdots \widehat{d^{c_k}}(z_k) \cdots d^{c_n}(z_n)
  A_{i_1}^{d_1}(t_1) \cdots  A_{i_m}^{d_m}(t_m)  } \es
+ \dsum{k=1}{m} \e_{ii_k}  \vev{  b^{d_k}(t_k) c^b(y)
  d^{c_1}(z_1) \cdots d^{c_n}(z_n)
  A_{i_1}^{d_1}(t_1) \cdots \widehat{A_{i_k}^{d_k}}(t_k)
                            \cdots  A_{i_m}^{d_m}(t_m)  } .
\ea\eqn{<ada>}
 They allow to compute all the Green functions of $A$ and $d$ from
the Green functions \equ{<bc>}, \equ{<bcdaa>} and \equ{<bcd>}
of the ghost fields determined by the antighost
equation. We thus obtain
\eq
\vev{ (A)^m (d)^n } = 0\qquad {\rm for}\quad m \ge 3 ,
\eqn{<aaad>}
and, for the nonvanishing, ghost independent, connected Green functions,
the relations
\eq\ba{l}
\vev{ A_i^b(y) d^{c_1}(z_1) \cdots d^{c_n}(z_n)  } \es
= g  \dsum{k,l=1 (l\not=k)}{n} f_{c_k c_l e}
 [\th(z_k^3-z_l^3)+\a^{(n)}] \pa_i \d^2(\zt_k-\zt_l)\es
\qquad \vev{  b^e(z_l) c^b(y)
  d^{c_1}(z_1) \cdots \widehat{d^{c_k}}(z_k)
  \cdots  \widehat{d^{c_l}}(z_l) \cdots d^{c_n}(z_n) }, \es
( n\ge2) ,
\ea\eqn{<add>}

\eq\ba{l}
\vev{ A_i^a(x) A_j^b(y) d^{c_1}(z_1) \cdots d^{c_n}(z_n)  } \es
= g \e_{ij} \dsum{k=1}{n} f_{b c_k e}
 [\th(y^3-z_k^3)+\a^{(n)}] \d^2(\yt-\zt_k)\es
\qquad \vev{  b^e(z_k) c^a(x)
  d^{c_1}(z_1) \cdots \widehat{d^{c_k}}(z_k) \cdots d^{c_n}(z_n) }, \es
(n\ge1) .
\ea\eqn{<aad>}
Here again, like for the two-point functions,
Bose symmetry for the field $A$ fixes the value of the
integration constants:
\eq
\a^{(n)} = -\half
\eqn{alpha-n}
It is then easy to see from the above
that the following recurrence relations hold:
\eq\ba{l}
\vev{ A_i^a(x) d^{c_1}(z_1) \cdots d^{c_n}(z_n)  } \es
= g \dsum{k=1}{n} f_{ac_ke} [\th(x^3-z_k^3)-\half]
                                               \d^2(\xt-\zt_k)\es
\qquad \vev{  A_i^e(z_k)
  d^{c_1}(z_1) \cdots \widehat{d^{c_k}}(z_k) \cdots d^{c_n}(z_n) }, \es
( n\ge2) ,
\ea\eqn{add-rec}

\eq\ba{l}
\vev{ A_i^a(x) A_j^b(y) d^{c_1}(z_1) \cdots d^{c_n}(z_n)  } \es
= g \dsum{k=1}{n} f_{ac_ke} [\th(x^3-z_k^3)-\half]
                                               \d^2(\xt-\zt_k)\es
\qquad \vev{  A_i^e(z_k) A_j^b(y)
  d^{c_1}(z_1) \cdots \widehat{d^{c_k}}(z_k) \cdots d^{c_n}(z_n) }, \es
( n\ge1) .
\ea\eqn{aad-rec}
They are illustrated together with the ghost recurrence
relation \equ{<bcd>} (with $\a^{(n)}=-1/2$)
in Figs. \ref{figbc} to \ref{figaa}. The thin lines correspond to the
two-point functions as depicted in Fig. \ref{figprop}. The three-point
vertices $(cbA_3)$ and $(AAA_3)$ correspond respectively to the
expressions $g f_{abc}\int d^3z\cdots$
and $g\e_{ij}f_{abc}\int d^3z\cdots$. A 'hat' on an argument means
its omission.
\begin{figure}
\vspace{80mm}
\includegraphics{csa1.ps}
\caption{Graphical representation of the two-point functions
\protect\equ{<bc>}, \protect\equ{<aa>}, \protect\equ{g-cond-exc} (with
$\a=-1/2$).}
\label{figprop}
\end{figure}
\begin{figure}
\vspace{80mm}
\includegraphics{csa2.ps}
\caption{Graphical representation of Eq.\protect\equ{<bcd>}. }
\label{figbc}
\end{figure}
\begin{figure}
\vspace{80mm}
\includegraphics{csa3.ps}
\caption{Graphical representation of Eq.\protect\equ{add-rec}.}
\label{figa}
\end{figure}
\begin{figure}
\vspace{80mm}
\includegraphics{csa4.ps}
\caption{Graphical representation of Eq.\protect\equ{aad-rec}.}
\label{figaa}
\end{figure}

We have thus obtained the general result that, in the axial gauge, the
supersymmetry completely fixes the connected Green functions, with the
notable exception of those involving the Lagrange multiplier field $d$
only.
\section{ Consequences of gauge invariance}
The gauge Ward identity \equ{gauge-wi} yields, for the connected Green
functions of the Lagrange multiplier field the equations
\eq\ba{l}
\pa_{x^3} \vev{d^a(x) d^{c_1}(z_1) \cdots d^{c_n}(z_n) }\es
=g\dsum{k=1}{n} f_{a c_k e}\d^3(x-z_k)
   \vev{d^e(z_k) d^{c_1}(z_1) \cdots \widehat{d^{c_k}}(z_k)\cdots
                       d^{c_n}(z_n)
}.
\ea\eqn{d-equations}
The solution which respects scale invariance vanishes identically:
\eq
\vev{ d^{c_1}(z_1) \cdots d^{c_n}(z_n) } = 0\quad\forall n.
\eqn{ddd}

This ends the demonstration that the Ward identities of the Chern-Simons
theory in the axial gauge determine uniquely all its Green functions.
However, beyond the gauge fixing condition and the supersymmetry Ward
identities -- together with the antighost equation  -- we
have used only a small part of the gauge Ward identities, namely the
equations \equ{d-equations} for the field $d$.
The remaining ones, which involve also the other fields, as well as the
ghost equation \equ{ghost-eq} have to be --
and indeed have been -- checked explicitly.
\section{Conclusion}
The main result of this study is that the Green functions of the
three-dimensional Chern-Simons
theory in the axial gauge follow as the -- unique --
solution of the Ward identities
defining the model, without reference to any action principle.

More
remarkable, they are all determined by the gauge condition and the Ward
identities of topological supersymmetry only, with
the exception of the Green
functions \equ{ddd} of the Lagrange multiplier field. It is merely in
order to fix the latters that the gauge Ward identity is effectively
needed.

However, notwithstanding the latter point, and looking at the solution
of the gauge condition and of the supersymmetry
Ward identities, we remark
that it consists of exactly the Green functions that one
would calculate -- with the principal value prescription for the free
propopagators~\cite{kummer} --
from the gauge fixed action \equ{action}, which is itself a
solution of the gauge Ward identity.
This is at best seen from the graphical representations (see Figs.
\ref{figbc} to \ref{figaa}) of the equations
\equ{<bcd>}, \equ{add-rec} and \equ{aad-rec}.
Turning the argument
round, we may conclude that enforcing the supersymmetric
Ward identities and the gauge condition on the Green functions fixes
uniquely the action -- solution of the gauge
Ward identity -- and then all the Green functions of the theory.
\section*{Appendices:}
Appendix A is devoted to demonstrate the existence of a
larger algebra of symmetries of the Chern Simons model quantized
in the axial gauge.\\ Parallel to the analysis done in the
covariant Landau gauge we show in App. B that also in the
non-covariant axial gauge an off-shell formulation for the BRS
and supersymmetry is possible.
\appendix
\section{The algebra generated by BRS, anti-BRS and
supersymmetry}\label{algebra}
Beside the symmetry transformations \equ{brs} and \equ{susy} one
can construct a further symmetry defined by
\eq\ba{l} \hat{\d}A_\m =- D_\m (2d- g\{b,c\}) , \es
   \hat{\d }b = g\left[d,b\right] , \es
   \hat{\d }c = g\left[d,c\right] , \es
   \hat{\d }d = 0.
   \ea\eqn{del-sym}
It is straightforward to show that the action \equ{action} is
invariant under \equ{del-sym} and in addition one has the
following closure with $s$ and $\bs$:

\eq\ba{ll} s^2=\bs^2 =0,&
   \{s,\bs\} = \hat{\d} , \es
   \left[ s , \hat{ \d } \right] = 0, & \left[ \bs, \hat{\d }\right] =0.
      \ea\eqn{del-alg}
In contradiction to the Landau gauge $s$ and $\bs$ do not
anticommute.\\
Nevertheless, $\n_\m$ and $\bv_\r$ form a closed algebra with
the ghost-number transformation $r$ and a new tensor symmetry
$h_{\a\b}$:
\eq\ba{ll} h_{\a\b}A_\m = -\e_{\a \m \n} n^\n A_\b -\e_{\b\m\n}n^\n A_\a,
&r A_\m =0 , \es
h_{\a\b}b= 0     , &rb= -b ,   \es
h_{\a\b}c= 0     ,  &rc= c ,   \es
h_{\a\b}d= -\pa_\a A_\b -\pa_\b A_\a          , &rd=0,
\ea\eqn{h-r-sym}
leading to the following larger algebra:
\eq\ba{ll} \{\n_\a,\n_\b\}=0  , &\{\bv_\a,\bv_\b\}=0 , \es
\{\n_\a,\bv_\b\}=h_{\a\b}+\e_{\a\b\n} n^\n r , & \es
\left[\n_\a, r\right]=\n_\a   , & \left[\bv_\a, r\right]=-\bv_\a , \es
\left[\n_\a,h_{\b\g} \right]=\e_{\b\a\n}n^\n \n_\g + \e_{\g\a\n}n^\n \n_\b , &
\left[\bv_\a,h_{\b\g} \right]=-\e_{\b\a\n}n^\n \bv_\g -
 \e_{\g\a\n}n^\n \bv_\b,\es
\left[r,h_{\a\b}\right] = 0.&
\ea\eqn{h-r-alg}
Additionally there exists a further symmetry algebra. The
two new symmetries $\tau_\r$ and $\tau$
\eq\ba{ll} \tau_\a A_\m =-\e_{\a\m\n} n^\n gc^2        , &\tau A_\m =0 , \es
   \tau_\a b = \pa_\a c +D_\a c                , &\tau b= gc^2 , \es
   \tau_\a c =0                                , &\tau c=0 , \es
   \tau_\a d=\pa_\a (g c^2)                     , &\tau d=0
   \ea\eqn{tau-sym}
lead to the following closed algebra:
\eq\ba{ll} \{s,\bv_\a\}=\tau_\a     , &\left[\tau_\a,\tau_\b\right]=0 , \es
   \left[\tau_\a,\bv_\b\right]=3\e_{\a\b\n} n^\n \tau    , &\tau^2=0 , \es
   \left[s,\tau_\a\right]=0        , &\{s,\tau\}=0 , \es
   \{\bv_\a,\tau\}=0     , &\left[\tau_\a,\tau\right]=0.
   \ea\eqn{tau-alg}
\section{Off-shell BRS and supersymmetry algebra}
As  usually and parallely to the analysis for the Landau gauge the
composite  fields appearing in the BRS transformations
\equ{brs} are coupled to external fields $\g^\m$,
$\s$, and thus induce the source term in the action:
\eq
\Sigma_{\rm ext}=\tr\dint d^3x \left(\g^\mu sA_\m +\s sc \right) .
\eqn{ext}
BRS invariance is then expressed by the Slavnov identity
\eq
\SS \zc = \tr\dint d^3x \left(
  J^\m \frac{\delta}{\delta\gamma^\m} - J_c \frac{\delta}{\delta\sigma}
  -J_b \dfud{}{J_d}   \right) \zc = 0,
\eqn{slavnov-ext}
where $\zc(J^\m,J_b,J_c,J_d,\g^\m,\s)$ is the generating functional of
the connected Green
functions, $J^\m$, $J_d$, $J_b$ and $J_c$ denoting the sources of the
fields $A_\m$, $d$, $b$ and $c$, respectively.

The presence of these source terms breaks the supersymmetry.
Thus one has the following Ward identity:
\eq\ba{rl}
\WW_\r \zc \equiv \tr\dint& d^3x \lp
 J^\m\e_{\r\m\n} n^\n\dfud{}{J_b}
-J_c\dfud{}{J^\r} + J_d\pa_\r \dfud{}{J_b}\right. \es
&\left. - \s \lp \dfud{}{\g^\r}-\pa_\r \dfud{}{J_c} \rp
- \g^\m \lp \pa_\r\dfud{}{J^\m} - \e_{\r\m\n}n^\n \dfud{}{J_d}
\rp\rp \zc \es
&=\D_\r= \tr\dint d^3x \e_{\r\m\n}J^\m\g^\n  .
\ea\eqn{susy-wi-ext}
The action of
the Slavnov operator on the ''breaking term'' $\D_\r$ vanishes
identically:
\eq\ba{rl}
\SS \D_\r&=0. \ea\eqn{brea-brs}
Remember the nilpotency  of the Slavnov operator
\eq \SS \SS F[J_\f]=0 . \eqn{nilpot}
for any functional $F[J_\f]$ of the sources, then the following
identities complete the algebra between supersymmetry and BRS.
\eq \lp\WW_\r \WW_\s + \WW_\s \WW_\r \rp F[J_\f] =0 . \eqn{ww-prod}
\eq\ba{rl} \WW_\r \SS F[J_\f] + \SS\lp\WW_\r F[J_\f]- \D_\r\rp
&=  \es\lp \WW_\r \SS +\SS \WW_\r \rp F[J_\f]&=\PP_\r F[J_\f].
\ea\eqn{ws-prod}
Here $\PP_\r$ is the operator of translations defined by
\eq\ba{rl} \PP_\r \zc \equiv \tr\dint d^3x &\lp \sum_\f \pa_\r
J_\f \dfud{}{J_\f}\rp \zc . \ea\eqn{translop}

\end{document}